\documentclass[preprint2]{aastex}

\usepackage{ulem}
\usepackage{bm}
\usepackage{amsmath}
\usepackage{lineno}
\shorttitle{Solar Wind Heating with Counter-propagating Waves} \shortauthors{He et al.}

\begin{document}

\title{Proton Heating in Solar Wind Compressible Turbulence with Collisions between Counter-Propagating Waves}

\author{Jiansen He\altaffilmark{1,2}, Chuanyi Tu\altaffilmark{1}, Eckart Marsch\altaffilmark{3}, Christopher H. K. Chen\altaffilmark{4}, Linghua Wang\altaffilmark{1}, Zhongtian Pei\altaffilmark{1}, Lei Zhang\altaffilmark{2}, Chadi S. Salem\altaffilmark{5}, Stuart D. Bale\altaffilmark{5,6}}

\altaffiltext{1}{School of Earth and Space Sciences, Peking University, Beijing,
100871, China; E-mail: jshept@gmail.com}
\altaffiltext{2}{Sate Key Laboratory of Space Weather, Chinese Academy of Sciences, Beijing 100190, China}
\altaffiltext{3}{Institute for Experimental and Applied Physics, Christian-Albrechts-Universit\"{a}t zu Kiel, 24118 Kiel, Germany}
\altaffiltext{4}{Department of Physics, Imperial College London, London SW7 2AZ, UK}
\altaffiltext{5}{Space Sciences Laboratory, University of California, Berkeley, California 94720, USA}
\altaffiltext{6}{Department of Physics, University of California, Berkeley, California 94720, USA}

\begin{abstract}
Magnetohydronamic turbulence is believed to play a crucial role in heating the laboratorial, space, and astrophysical plasmas. However, the precise connection between the turbulent fluctuations and the particle kinetics has not yet been established. Here we present clear evidence of plasma turbulence heating based on diagnosed wave features and proton velocity distributions from solar wind measurements by the Wind spacecraft. For the first time, we can report the simultaneous observation of counter-propagating magnetohydrodynamic waves in the solar wind turbulence. Different from the traditional paradigm with counter-propagating Alfv\'en waves, anti-sunward Alfv\'en waves (AWs) are encountered by sunward slow magnetosonic waves (SMWs) in this new type of solar wind compressible turbulence. The counter-propagating AWs and SWs correspond respectively to the dominant and sub-dominant populations of the imbalanced Els\"asser variables. Nonlinear interactions between the AWs and SMWs are inferred from the non-orthogonality between the possible oscillation direction of one wave and the possible propagation direction of the other. The associated protons are revealed to exhibit bi-directional asymmetric beams in their velocity distributions: sunward beams appearing in short and narrow patterns and anti-sunward broad extended tails. It is suggested that multiple types of wave-particle interactions, i.e., cyclotron and Landau resonances with AWs and SMWs at kinetic scales, are taking place to jointly heat the protons perpendicularly and parallel.
\end{abstract}


\section{Introduction}

Turbulence is a common phenomenon in various environments, e.g., laboratorial \citep{Howes2012PhRvL.109y5001H}, space \citep{Tu1995SSRv, Goldstein1995ARA&A, Bruno2013LRSP}, and astrophysical plasmas \citep{Zhuravleva2014Nature}. In the heliosphere the solar wind is deemed to be an ideal natural laboratory to investigate plasma turbulence, since particles and fields can be measured in-situ and the solar wind is approximately boundless for the turbulent eddies/fluctuations therein. The magnetic fluctuations are often correlated with the velocity fluctuations. Together with the weak compressibility of the magnetic field strength, this indicates the dominance of Alfv\'enic fluctuations in the turbulence \citep{Belcher1971JGR}. The so-called Els\"asser variables, which are defined as a sum or difference of fluctuations for velocity and magnetic field in Alfv\'en unit ($\mathbf{Z}_{\pm}=\mathbf{v}\pm\mathbf{b}$), have often been employed in theoretical calculations to study the counter-propagating waves in Alfv\'enic turbulence. From the space observation point of view, the nature of the minor (subdominant) Els\"asser variable remains unknown, although it is assumed to represent sunward/inward Alfv\'en waves in the scenario of Alfv\'enic turbulence. However, this notion has not yet been confirmed. Nevertheless, sunward propagating Alfv\'en waves with the absence of anti-sunward Alfv\'en waves are identified in upstream regions of the bow shock \citep{Wang2015GeoRL} and in some streams of the pristine solar wind \citep{He2015ApJ-sunward}.

The minor Els\"asser field may also be associated with convected magnetic structures \citep{Tu1993JGR, Bruno2001P&SS}, which would also contribute to the observed excess of magnetic energy over kinetic energy \citep{Chen2013ApJ}. Therefore, it is still unclear whether $\mathbf{Z}_{\rm{minor}}$ represents inward waves and/or involves mostly nonlinear interaction with the outward $\mathbf{Z}_{\rm{major}}$. The compressibility of solar wind turbulence usually manifests itself in pressure-balanced structures of various types \citep{Yao2011ApJ}, which may come in the form of spaghetti-like flux tube stemming from the Sun \citep{Borovsky2008JGRA}, quasi-perpendicular kinetic slow magnetosonic waves \citep{Howes2012ApJ, Zhang2015AnGeo}, or mirror-mode structures \citep{Zhang2009JGR, Yao2013ApJ}. The nature of $\mathbf{Z}_{\rm{minor}}$ and its relation to compressibility will be investigated here.

Turbulence is thought to be dissipated through the damping of small-scale kinetic waves or via currents in coherent structures, and can thereby heat the plasma through energy conversion \citep[e.g.,][]{Voitenko2004NPGeo, Marsch2006LRSP, Markovskii2011ApJ}. For example, both linear and nonlinear resonances between waves and particles can break the adiabatic behavior of particles \citep{Marsch1982JGR, Hollweg2002JGRA}. Resonance diffusion plateaus of various types, e.g., Landau resonance and left/right cyclotron resonance, seem to be consistent with the measured contours of proton velocity distributions in fast solar wind streams \citep{Marsch2001JGR, He2015ApJ}, where turbulence is dominated by Alfv\'en waves from the Sun. Non-resonant (stochastic) heating might be important in regions with extremely low plasma beta, e.g., close to the solar corona \citep{Chandran2010ApJ}. Dissipation of strong current density within coherent structures may also be an important source of local heating \citep{Parashar2009PhPl, Osman2011ApJ, Karimabadi2013PhPl, TenBarge2013ApJ, Wang2013ApJ, Zhang2015ApJ}. But how particles are heated in compressible MHD turbulence involving counter-propagating waves still remains to be understood. This question will also be studied here in detail by use of Wind in-situ particle and field measurements.

\section{Observations and data analysis}

The solar wind observations presented here are from the Wind spacecraft, and are based on the plasma and magnetic field data provided respectively by the 3DP and MFI instruments \citep{Lin1995SSRv, Lepping1995SSRv}. The time interval for our study is [15:00, 18:00] on 2012/12/31. Time profiles of parameters, i.e., proton number density ($N_{\rm{P}}$), bulk velocity components ($V_\parallel$, $V_{\perp 1}$, and $V_{\perp 2}$), magnetic field components ($B_\parallel$, $B_{\perp 1}$, and $B_{\perp 2}$), are illustrated in the left panels of Figure~1. The parallel and the first and second perpendicular directions are defined as the directions of $\mathbf{b}_{\rm{0}}$, $\mathbf{b}_{\rm{0}}\times \mathbf{x}$ and $\mathbf{b}_{\rm{0}} \times (\mathbf{b}_{\rm{0}} \times \mathbf{x})$, where $\mathbf{b}_{\rm{0}}$ represents the mean magnetic field direction pointing anti-sunward in this case, and $\mathbf{x}$ is the GSE x-direction. It is surprising to find that: $V_\parallel$ is positively correlated with $B_\parallel$, indicating sunward SMWs propagating obliquely to $\mathbf{b}_{\rm{0}}$; $V_{\perp 1}$ and $V_{\perp 2}$ are anti-correlated respectively with $B_{\perp 1}$ and $B_{\perp 2}$, indicating anti-sunward propagating AWs. The finding of sunward propagating SMWs is also confirmed by the anti-correlation between $N_{\rm{P}}$ and $V_\parallel$.

The right panels of Figure~1 illustrate the cross-correlation spectra calculated by means of the formula, ${\rm{CC}}_{{ab}}  = {\mathop{\rm Re}\nolimits} (\tilde W_{{a}} \tilde W_{{b}}^* )/(\left| {\tilde W_{{a}} } \right| \cdot \left| {\tilde W_{{b}} } \right|)$, where $\tilde W _{{a}}$ and $\tilde W_{{b}}$ are the wavelet coefficient spectra of $a$ and $b$. Spectra of negative ${\rm{CC}}(N_{\rm{P}},\left|\mathbf{B}\right|)$, ${\rm{CC}}(N_{\rm{P}},V_\parallel)$, and positive ${\rm{CC}}(V_\parallel,B_\parallel)$ are clear evidence of sunward oblique SMWs. For oblique SMWs, the density fluctuation is driven by longitudinal compression/expansion, and the magnetic field pressure is almost in balance with the thermal pressure. The anti-sunward AWs are characterized by spectra having negative ${\rm{CC}}(V_\perp,B_\perp)$. The evidence of counter-propagating waves is solid, and not affected by the definition of $\mathbf{b}_{\rm{0}}$, whether it is a global mean field averaged over the whole time interval (Figure~1b-1e) or a local mean magnetic field direction as introduced by \citet{Horbury2008PhRvL} (Figure~1g-1j).

The Els\"asser variables $\mathbf{Z}_{\pm}$ are calculated by $\mathbf{Z}_{\pm}=\mathbf{V}\pm \mathbf{V}_{\rm{a}}=\mathbf{V}\pm \mathbf{B}/\sqrt {\mu_{\rm{0}} \rho_{\rm{0}} }$, with $\rho_{\rm{0}}$ being the proton density averaged over the interval of [15:00, 18:00] UT. The trace power spectral densities (PSD) of $\mathbf{Z}_{\rm{+}}$, $\mathbf{Z}_{\rm{-}}$, $\mathbf{V}$, and $\mathbf{V}_{\rm{a}}$ are illustrated in Figure~2a and 2b.  The related spectral indices are found by power-law fitting to be $-1.55\pm0.04$, $-1.54\pm0.03$, $-1.51\pm0.03$, $-1.52\pm0.03$.

The characteristic directions of $\mathbf{Z}_{\pm}$ fluctuations are estimated by applying the Singular-Value-Decomposition method \citep{Santolik2003RaSc} to the spectral matrix of $\mathbf{Z}_{\pm}$. The directions with maximum singular values are regarded as the oscillation directions. The direction with minimum singular value is deemed a good proxy for the propagation direction, if the Els\"asser fluctuations are ideally incompressible. The spectra of the angles between the derived ``oscillation'' or ``propagation'' direction and the local mean magnetic field directions ($\theta(\mathbf{Z}_{\rm{+}},\mathbf{b}_{\rm{0}})$, $\theta(\mathbf{Z}_{\rm{-}},\mathbf{b}_{\rm{0}})$, $\theta(\mathbf{k}_{\rm{+}}^*,\mathbf{b}_{\rm{0}})$, $\theta(\mathbf{k}_{\rm{-}}^*,\mathbf{b}_{\rm{0}})$) as a function of time and period are illustrated in Figure~2c, 2d, 2e, 2f. The angle $\theta(\mathbf{Z}_{\rm{+}},\mathbf{b}_{\rm{0}})$ in Figure~2c is basically smaller than $30^\circ$, indicating longitudinal oscillation of $\mathbf{Z}_{\rm{+}}$. Large angles $\theta(\mathbf{Z}_{\rm{-}},\mathbf{b}_{\rm{0}})$ in Figure~2d suggest a transverse oscillation of $\mathbf{Z}_{\rm{-}}$.

The oscillation difference between $\mathbf{Z}_{\rm{+}}$ and $\mathbf{Z}_{\rm{-}}$ demonstrates the intrinsically different natures of $\mathbf{Z}_{\rm{+}}$ and $\mathbf{Z}_{\rm{-}}$, with the former and latter pertaining to the sunward SMWs and anti-sunward AWs, respectively. Figure~2e shows that the majority of $\theta(\mathbf{k}_{\rm{+}}^*,\mathbf{b}_{\rm{0}})$ is larger than $70^\circ$, while $\theta(\mathbf{k}_{\rm{-}^*,\mathbf{b}_{\rm{0}}})$ in Figure~2f displays a scattered probability distribution. To be cautious, $\theta(\mathbf{k}_{\rm{+}}^*,\mathbf{b}_{\rm{0}})$ may not be the real propagation angle of SMWs due to the compressibility of $\mathbf{Z}_{\rm{+}}$. The derived $\mathbf{k}_{\rm{-}}^*$ for Alfv\'en waves may be either the real propagation direction $\mathbf{k}_{\rm{-}}$ or $\mathbf{b}_{\rm{0}}$. The patches of $\theta(\mathbf{k}_{\rm{+}}^*,\mathbf{Z}_{\rm{-}})\neq 90^\circ$ in Figure~2g may infer the nonlinear action of AWs ($\mathbf{Z}_{\rm{-}}$) on SMWs ($\mathbf{Z}_{\rm{+}}$) ($\mathbf{Z}_{\rm{-}} \cdot \nabla_{\rm{+}} \mathbf{Z}_{\rm{+}} \neq 0$). The AWs may be shorn by the SMWs when $\theta(\mathbf{k}_{\rm{-}}^*, \mathbf{Z}_{\rm{+}}) \neq 90^\circ$ (Figure~2h), which may lead to a transverse cascade of AWs.

The proton 2D velocity distribution functions (VDFs) in the top panels of Figure~3 illustrate the bi-directional asymmetric beams along the $\mathbf{b}_{\rm{0}}$ direction. Such a twofold beam pattern has, to our knowledge, not been reported before. As compared to the anti-sunward beam, which appears prolonged in $V_\parallel$ and broad in $V_\perp$, the sunward beam looks relatively shorter in $V_\parallel$ and narrower in $V_\perp$. We interpret this new observational pattern as probable evidence of wave-particle interaction between protons and counter-propagating waves.

The anti-sunward beam may be the result of Landau and right-cyclotron resonance with the outward oblique AWs at kinetic scales (generalized kinetic Alfv\'en waves), the wave signatures of which had been reported in previous studies \citep{Bale2005PhRvL, Sahraoui2010PhRvL, He2011ApJ, He2012ApJ, Podesta2011ApJ, Roberts2013ApJ, Klein2014ApJ}. The sunward beam may be formed via Landau resonance with kinetic SMWs, the wave features of which have been studied in linear theory \citep{Zhao2014ApJ, Narita2015ApJ}. In other time intervals, without clear signatures of counter-propagating MHD waves, the bi-directional asymmetric beams in the VDFs are absent. The proton 1D reduced VDFs in the bottom panels of Figure~3 show a plateau and an extended tail on the sunward and anti-sunward sides respectively. It is exciting to find that the plateau positions coincide well with the field-projected phase speed of quasi-perpendicular SMWs (vertical dotted line in panels), which indicates a Landau resonance between protons and sunward SMWs. The dispersion relation for quasi-perpendicular SMW yields approximately ${C_{\rm{S}}^*}^2=({\omega }/{{k_\parallel}})^2 \sim {{V_{\rm{A}}^2 }}/({{1 + V_{\rm{A}}^2 /C_{\rm{S}}^2}})$, with $V_{\rm{A}}$ and $C_{\rm{S}}$ being the Alfv\'en speed and ion sound speed respectively. The propagation of quasi-perpendicular SMWs can also be described by the equations derived for the generalized Els\"asser fields \citep{Marsch1987JGR}. But they have never before been analyzed observationally. SMWs in the extreme limit of $k_\parallel \ll k_\perp$ are also named as pseudo-Alfv\'enic waves, which is passive scattered/mixed by Alfv\'en waves with $k_\parallel \ll k_\perp$ \citep{Schekochihin2009ApJS}. Alfv\'en waves with considerable $k_\parallel$ may also be distorted by the quasi-perpendicular SMWs leading to larger $k_\perp$ due to the nonlinear term $Z_{\parallel,{\rm{SMW}}}\cdot \nabla Z_{\perp,{\rm{AW}}}$ in the convective derivative.

\section{Summary and discussions}

On the basis of the observational findings presented here, a new scenario of compressible plasma turbulence is proposed and illustrated in Figure~4. Dominant AWs are propagating anti-sunward and quasi-perpendicular SMWs are propagating upstream against the AWs. The nonlinear interaction during collisions between these counter-propagating waves may lead for both waves to a cascade of wave energy towards plasma kinetic scales. Meanwhile, in response to the counter-propagating waves, bi-directional asymmetric beams emerge in the proton VDFs through resonant wave-particle interactions. In this way, the turbulence energy cascading and dissipation as well as the solar wind ion heating are included in this common scenario. This new type of turbulence, which is unusual in the solar wind fast streams, is different from the traditional paradigm, where counter-propagating waves are both Alfv\'enic but not yet confirmed observationally.

Some new challenges are raised along with these findings. (1) How are the sunward SMWs generated in the solar wind turbulence: Is it by compression between streams, the parametric decay from AWs \citep{Verscharen2012PhPl}, or by other processes? (2) The observed compressible MHD turbulence and the concurrent collisionless damping of kinetic waves, together call for a more comprehensive study, involving numerical modeling and observational statistical analysis. (3) What is the radial evolution in the solar wind of the compressible turbulence, involving AWs and SMWs, which have also been identified in the solar wind source region? (4) Besides the local wave-particle interactions, effects of large-scale evolution (e.g., anisotropic cooling during expansion, gravity, charge-separation electric field, mirror force, and differential flows between species) need to be taken into account for the proton kinetics \citep[e.g.,][]{Cranmer2014ApJS, Isenberg2015ApJ, Hellinger2015JPlPh}. These are issues that may well be addressed by the upcoming missions Solar Orbiter and Solar Probe Plus, which will be "sailing against the solar wind" into the inner heliosphere and outer corona.

\begin{acknowledgements}
{\bf Acknowledgements:} This work at Peking University is supported by NSFC under contracts 41222032, 41174148, 41574168, 41231069, 41274172, 41474148, and 41421003. The work at UC Berkeley is supported by NASA grants NNX13AL11G and NNX14AC07G. JSH, CYT, EM, CHKC, and LHW are members of ISSI/ISSI-BJ Team 304. We thank the 3DP team and MFI team on WIND for sharing data with us.
\end{acknowledgements}

\begin{figure*}
\centering
\includegraphics[width=16cm]{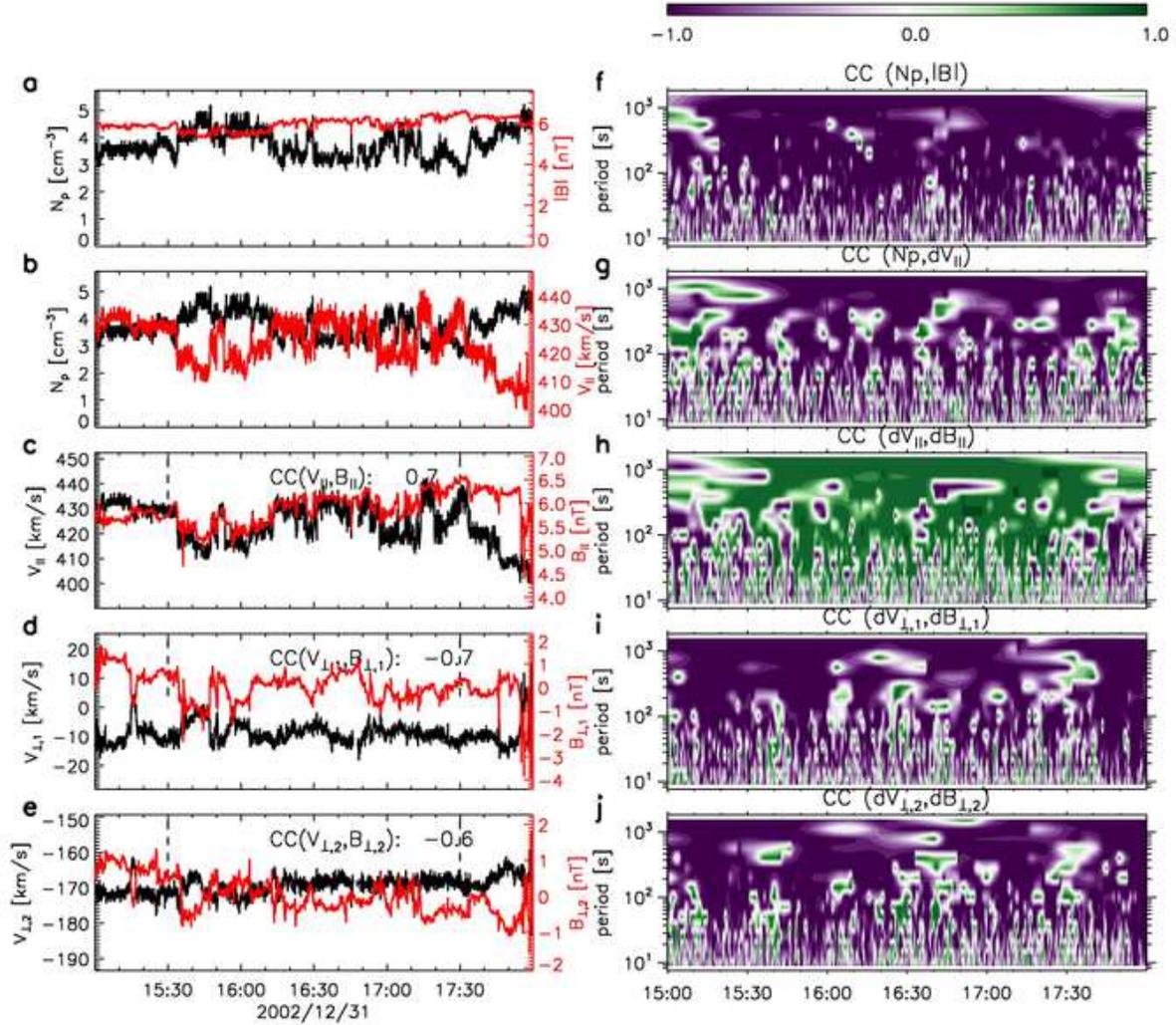}
\caption{Evidence of sunward quasi-perpendicular slow magnetosonic waves (SMWs) coexisting with anti-sunward Alfv\'en waves (AWs); the evidence is based on the correlation analysis of various parameter pairs. Time profiles of $N_{\rm{p}}$ (black) and $|\mathbf{B}|$ (red) in (a), $N_{\rm{p}}$ (black) and $V_\parallel$ (red) in (b), $V_\parallel$ (black) and $B_\parallel$ (red) in (c), $V_{\perp,1}$ (black) and $B_{\perp,1}$ (red) in (d), $V_{\perp,2}$ (black) and $B_{\perp,2}$ (red) in (e). Correlation spectra between different pairs of parameters ($N_{\rm{p}}$, $|\mathbf{B}|$) in panel (f), ($N_{\rm{p}}$, $V_\parallel$) in (g), ($V_\parallel$, $B_\parallel$) in (h), ($V_{\perp,1}$, $B_{\perp,1}$) in (i), and ($V_{\perp,2}$, $B_{\perp,2}$) in (j). The indices $\perp,1$ and $\perp,2$ refer to the directions of $\mathbf{b}_{\rm{0}}\times\mathbf{x}$ and $\mathbf{b}_{\rm{0}} \times(\mathbf{b}_{\rm{0}}\times\mathbf{x})$. The direction $\mathbf{x}$ represents the GSE-x direction. The direction $\mathbf{b}_{\rm{0}}$ in the left and right panels refers to the global and local mean magnetic field direction, respectively.} \label{Fig.1}
\end{figure*}

\begin{figure*}
\centering
\includegraphics[width=16cm]{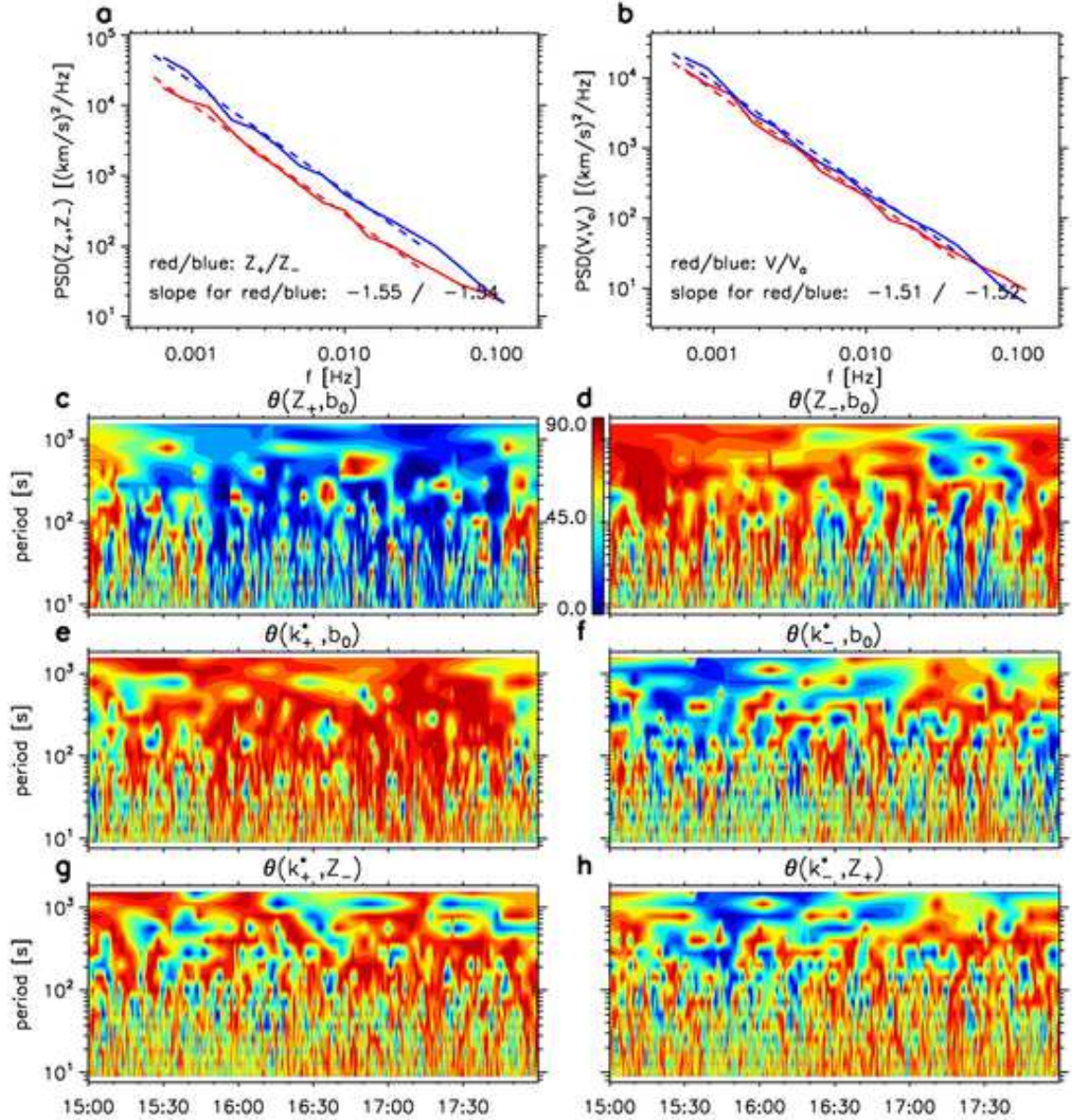}
\caption{The Els\"asser variables reveal the fluctuations to be longitudinal oscillations of sunward quasi-perpendicular SMWs for $\mathbf{Z}_{\rm{+}}$ and transverse oscillations of anti-sunward AWs for $\mathbf{Z}_{\rm{-}}$. Trace power spectra of $\mathbf{Z}_{\rm{+}}$ \& $\mathbf{Z}_{\rm{-}}$ in (a), and $\mathbf{V}$ \& $\mathbf{V}_{\rm{a}}$ in (b) during the time interval, with their fitted spectral indices being $-1.55\pm0.04$, $-1.54\pm0.03$, $-1.51\pm0.03$ and $-1.52\pm0.03$, respectively. (c) Spectrum of angle between $\mathbf{Z}_{\rm{+}}$ and $\mathbf{b}_{\rm{0}}$, where $\mathbf{Z}_{\rm{+}}$ direction is the characteristic direction with maximum singular values as obtained from Singular Value Decomposition method, and $\mathbf{b}_{\rm{0}}$ direction is the local mean magnetic field direction. Small angles coded in blue indicate the longitudinal oscillation of $\mathbf{Z}_{\rm{+}}$ along $\mathbf{b}_{\rm{0}}$. (d) Spectrum of $\theta(\mathbf{Z}_{\rm{-}},\mathbf{b}_{\rm{0}})$ with large angles in red showing oscillation of $\mathbf{Z}_{\rm{-}}$ transverse to $\mathbf{b}_{\rm{0}}$. (e) Spectrum of $ \theta(\mathbf{k}_{\rm{+}}^*,\mathbf{b}_{\rm{0}})$, where $\mathbf{k}_{\rm{+}}^*$ direction that is basically representing the eigenvector with minimum variance of $\mathbf{Z}_{\rm{+}}$ may have some deviation from the real propagation direction due to the compressibility. (f) Spectrum of $\theta(\mathbf{k}_{\rm{-}}^*,\mathbf{b}_{\rm{0}})$ with $\mathbf{k}_{\rm{-}}^*$ being the characteristic direction of $\mathbf{Z}_{\rm{-}}$ minimum variance. The possible nonlinear impact of $\mathbf{Z}_{\rm{-}}$ on $\mathbf{Z}_{\rm{+}}$ may be inferred from the patches with $ \theta(\mathbf{k}_{\rm{+}}^*,\mathbf{Z}_{\rm{-}})<90^\circ$ in (g). Patches with small angles in (h) imply the possible role of SMWs in cascading the energy of outward AWs.} \label{Fig.2}
\end{figure*}

\begin{figure*}
\centering
\includegraphics[width=16cm]{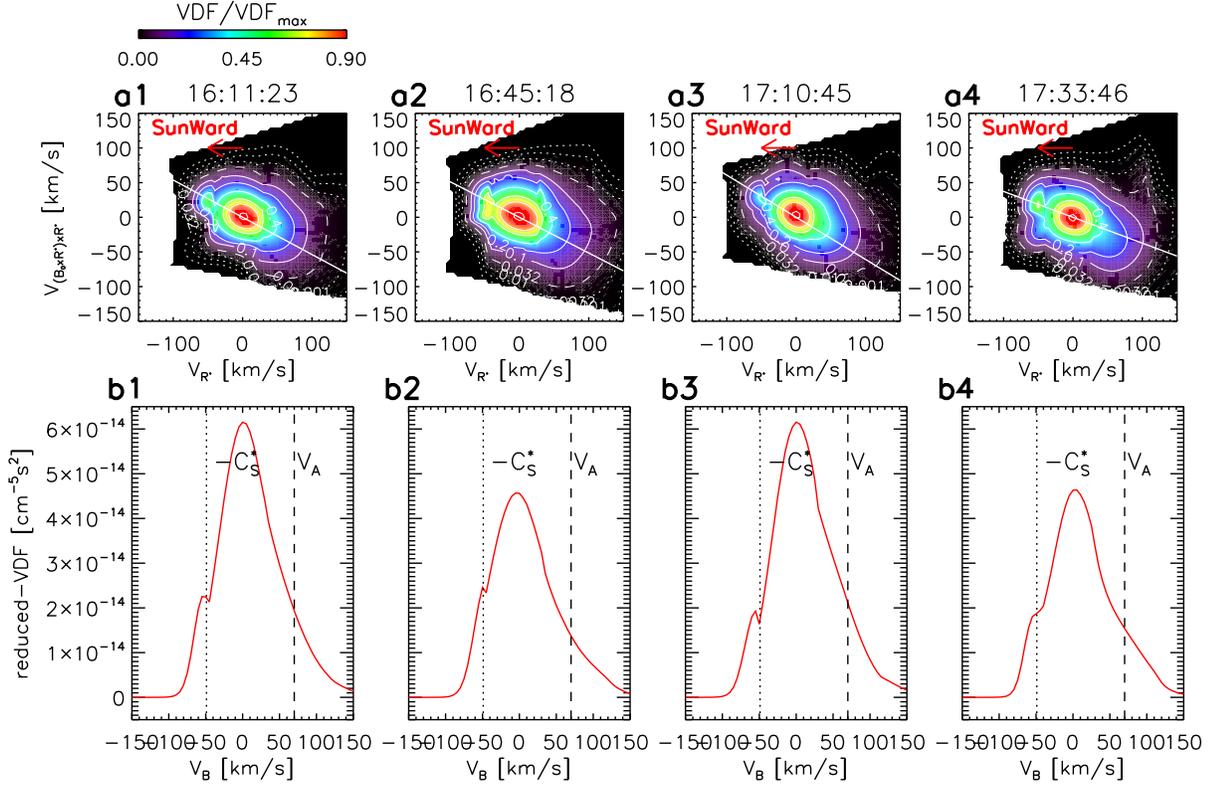}
\caption{Bi-directional asymmetric proton beams associated with the counter-propagating waves, indicating the solar wind heating by turbulence via Landau and cyclotron resonances. (a1) Cut view of proton velocity distributions at 16:11:23 with the GSE-x direction ($\mathbf{R}^*$ defined as -x in GSE) and magnetic field direction ($\mathbf{B}_{\rm{0}}$) in the plane, showing sunward and anti-sunward beams on either side of the core are aligned with the magnetic field direction ($\mathbf{B}_{\rm{0}}$, white line). Different from the anti-sunward beam, which extends further from the core and has a broad profile across $\mathbf{B}_{\rm{0}}$, the sunward beam drifts at a smaller speed and appears to have a narrower transverse velocity profile. (b1) Reduced VDF as integrated from the VDF in the plane of ($\mathbf{B}_{\rm{0}}$, $\mathbf{R}^*\times\mathbf{B}_{\rm{0}}$) along the direction of $\mathbf{R}^*\times\mathbf{B}_{\rm{0}}$. Left dotted line indicate the position of $-C_{\rm{S}}^*$ (the phase speed of quasi-perpendicular slow mode wave as projected along $\mathbf{B}_{\rm{0}}$), and right dashed line is located at the position of $V_{\rm{A}}$ (Alfv\'en speed). Velocity distributions at another three times are illustrated in (a2, b2), (a3, b3), and (a4, b4).} \label{Fig.3}
\end{figure*}

\begin{figure*}
\centering
\includegraphics[width=16cm]{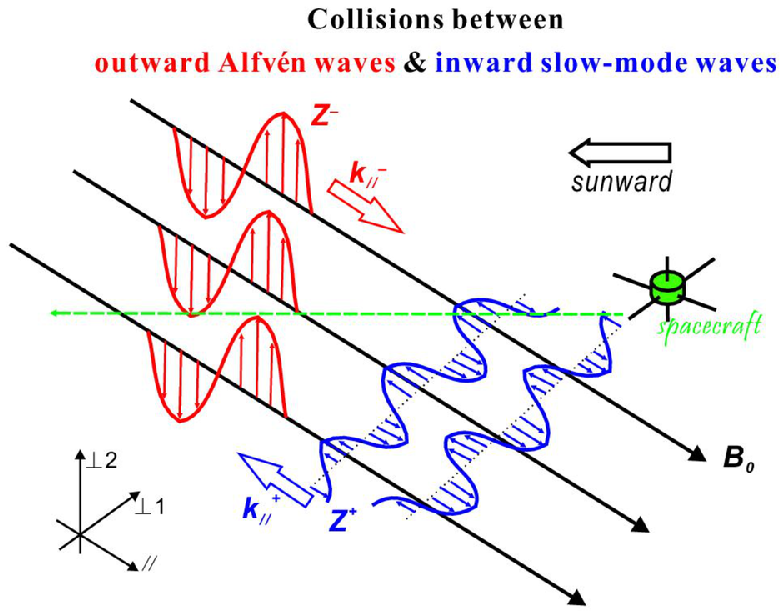}
\caption{The sketch illustrates solar wind turbulence that is characterized by collisions between outward AWs and inward SMWs. These collisions may lead to wave coupling, energy cascading, dissipation and heating. Transverse oscillations of $\mathbf{Z}_{\rm{-}}$ for outward AWs are described with three wave forms along three background magnetic field lines. Longitudinal oscillation of $\mathbf{Z}_{\rm{+}}$ for inward SMWs are represented by two wave forms aligned perpendicular to $\mathbf{B}_{\rm{0}}$. The trajectory of the WIND spacecraft in the solar wind frame is denoted by the green dashed line.} \label{Fig.4}
\end{figure*}

%\newpage
\bibliographystyle{apj}
%\bibliography{references}

\end{document}